# Core Services in the Architecture of the National Digital Library for Science Education (NSDL)

**Carl Lagoze (ed.)**
Computer Science Department
Cornell University
Ithaca, NY 14853 USA
+1 607 255 6046
lagoze@cs.cornell.edu[1]

**Walter Hoehn, David Millman**
Academic Information Systems
Columbia University
NY, NY 10025 USA

**William Arms, Stoney Gan, Diane Hillmann, Christopher Ingram, Dean Krafft, Richard Marisa, Jon Phipps, John Saylor, Carol Terrizzi**
Cornell University
Ithaca, NY 14853 USA

**James Allan, Sergio Guzman-Lara, Tom Kalt**
Center for Intelligent Information Retrieval
Department of Computer Science
University of Massachusetts
Amherst, MA 01003 USA

**ABSTRACT**
We describe the core components of the architecture for the (NSDL) National Science, Mathematics, Engineering, and Technology Education Digital Library. Over time the NSDL will include heterogeneous users, content, and services. To accommodate this, a design for a technical and organization infrastructure has been formulated based on the notion of a *spectrum of interoperability*. This paper describes the first phase of the interoperability infrastructure including the metadata repository, search and discovery services, rights management services, and user interface portal facilities.

**INTRODUCTION**

The NSDL (National Science, Mathematics, Engineering and Technology Education Digital Library) will possibly be the largest digital library ever constructed. Over the next five years the library is expected to serve millions of users and provide access to tens of millions of digital resources. A number of earlier articles [27, 32] describe the roots, vision, and strategy of the NSDL.

The National Science Foundation (NSF), through the Division of Undergraduate Education, is funding the initial development of the NSDL. Currently this funding amounts to about $24 million distributed across 64 projects. The majority of these projects are developing and contributing collections in various genre and subject areas. Others are building services that will enrich users' experience with these collections. Finally, the Core Integration (CI) funding, distributed among a group of collaborating institutions[1], is aimed at the development, deployment, and support of both the technical and organization infrastructure of the NSDL.

On their own these NSF-funded participants form a diverse group both in terms of the resources they bring to the library and their level of expertise with digital library technology. Yet, as the library develops, these projects will form only the nucleus of a much larger NSDL community. Some members of this larger community will participate as users, some will actively contribute related content, and some will become part of the library indirectly because they make available relevant content on the open access web.

The result will be a heterogeneous community of participants and technologies that will test many of the notions of *interoperability* that the digital library community has been investigating over the past decade. Digital library interoperability has many dimensions [22] and has been the subject of many initiatives including metadata [2], search and discovery [19, 21], resource naming [6], and service architecture [17].

Rather than prescribe a single approach for interoperability, the NSDL will accommodate heterogeneous parti-cipants, content, and users through a *spectrum of interoperability*. The motivations and concepts underlying this are discussed in some detail in a companion paper [10] that describes an earlier prototype of the NSDL base architecture. Stated briefly, this notion is based on a framing of the interoperability issue that plots the cost of adoption of an interoperability approach versus its functionality. Generally speaking, however great the functionality, an approach with high cost of adoption will not be widely used. To achieve widespread adoption, the cost of adoption must be low. This has been demonstrated many times over the past few decades, notably by the rapid and broad acceptance of basic web components such as HTTP and HTML in contrast to the limited distribution of more complex and rich mechanisms such as SGML and Z39.50.

---

[1] In addition to the institutions represented in the authorship of this paper, the CI team includes the University Center for Atmospheric Research (UCAR), the San Diego Supercomputer Center, the University of California at Santa Barbara.

The intent here is not to pass judgment on the complexity or functionality of any particular interoperability approach. Rather it is to promote the fact that different mechanisms are suited for different situations and implementation environments. In order for the NSDL to succeed as a library with breadth and depth it must embrace many points on the interoperability cost/functionality curve and provide participants with the tools to become involved at various points in the spectrum that are appropriate for their needs and capabilities.

Developing such an infrastructure is an ambitious task that must be approached in stages. Members of the CI technical team are currently implementing the first phase of this interoperability fabric. (In parallel, members of the team are developing plans on how to coordinate this base architecture with higher level services [13].) The goal of this phase is to develop the core services to facilitate the rapid growth of the library and incorporate services and content from both the NSDL funded contributors and those representing the broader web community. Following the notion that broad acceptance requires low entry cost, the architectural design for this phase is based on sharing of human and machine-generated metadata and exploitation of that metadata for the deployment of core services (e.g., search and discovery).

The components of the base architecture described in this paper are explicitly NOT the *final product* of the NSDL technical effort. Since low cost of entry correlates with limitations in functionality, NSDL's initial release will necessarily lack some of the features and extended services of more complex frameworks. As we all learned as children, however, it is important to walk before you run. Putting these initial walking steps in place will hopefully:

- Deploy as quickly as possible a publicly accessible digital library rich in science, technology, engineering, and mathematics resources.
- Build an experienced community of implementers of digital library technology and methods.
- Provide the necessary breathing room to understand and formulate the organizational aspects of running a large heterogeneous digital library
- Build a community that can work together to understand and develop future technical choices for the NSDL and make appropriate choices along the interoperability spectrum.

The architecture described in this paper is a realizable vehicle for reaching these goals.

## OVERVIEW OF THE ARCHITECTURE

Figure 1 illustrates the core components of the NSDL architecture and the interactions among them. The architecture builds on a number of fundamental concepts developed over the past several years of digital library research including:

- A common core metadata vocabulary to facilitate resource discovery across heterogeneous objects [28, 29].
- The integration of this core metadata with richer domain specific metadata [15].
- Harvesting of this metadata and its use as the basis for searching across repository contents and creation of richer services [11, 16].
- Automated indexing and retrieval systems that reduce the need for expensive cataloging [24].

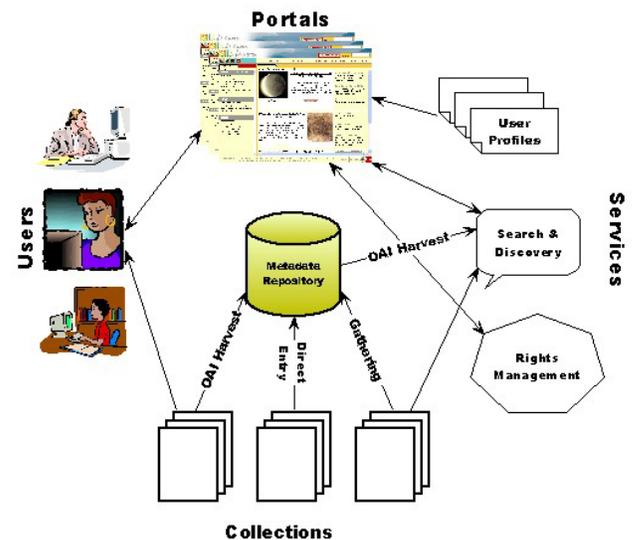

**Figure 1 - Components of NSDL Core Architecture**

As indicated in the figure, a key component of the architecture is the *metadata repository* (MR for the remainder of this paper) that holds metadata, in both common core and richer native forms, about all the NSDL collections and items in those collections. The contents of the MR either come directly from contributing content repositories or are derived computationally from the contents of content repositories. By combining metadata from many collections, the MR can be considered as a generalization of the concept of a union catalog, i.e., a catalog that combines records from many libraries.

The MR exposes the stored metadata through a set of interfaces for consumption by external services. These services process the metadata for targeted functions. As indicated in the figure, one core service is the search and discovery service that indexes metadata and associated content and presents a query API. The open modularity of the architecture makes it possible for this service set to grow over time.

It is important to note that users do not interact directly with the MR. Rather they use portals to access services that have processed the metadata. They then access the contents of collections directly, presumably guided to appropriate resources with the aid of the services. NSDL portals can be either general or audience specific (e.g., for teachers, students, researchers, etc.).

Access to the contents of the MR and the content described by that metadata is mediated by an access management service. This service is charged with both determining the identity of users and the groups they belong to and matching that group identity with appropriate access restrictions based on information in the metadata for the resource.

**METADATA REPOSITORY**

The MR provides a facility for storing metadata entities and the relationships between those entities, creating a platform on which to build essential services. It provides the following basic functions:

- Robust, central storage of metadata provided by NSDL participating partners as well as metadata gathered from non-participating open-access web resources.
- Output interfaces that provide metadata to services, such as search and browse services, that add value to the NSDL.
- Input interfaces that enable ingest services to provide new metadata and update existing metadata.

**NSDL Metadata Strategy**

The earlier Site for Science prototype [10] recognized that few of the collections selected for inclusion in the NSDL have metadata conforming to common or well-established standards, if they have metadata at all. Others have limited or idiosyncratic metadata, support other standards, or have local variations that cannot be handled easily.

Rather than deal with this reality by supporting a labor-intensive and costly central cataloging operation the NSDL is adopting the following strategy:

- Collect (through a variety of ingest mechanisms described later) item metadata from cooperating collections in any of eight supported "native" formats [20].
- When appropriate, automatically crosswalk native metadata to qualified Dublin Core [3], which will provide a lingua franca for interoperability.
- When item-level metadata does not exist and where possible, process content and generate metadata automatically [18].
- Accept that item-level metadata will not always exist but mandate that collection-level metadata always exists. Concentrate limited human effort on the creation of this collection-level metadata.
- Assemble all metadata, core and native, item and collection, in the central MR.
- Expose metadata records in the repository for service providers to harvest

**Repository Contents**

The MR contains metadata records. Metadata records take several forms as illustrated in Figure 2 and described below.

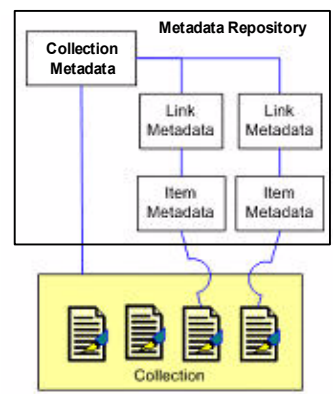

Figure 2 - NSDL Repository Records

- *Item metadata records* describe individual digital resources outside of the repository and function as surrogates for these resources. In this context, "resource" is a very general term, which can include: web pages, images, sounds, video clips, complete web sites, online or offline collections or groups of resources; online services, processes, databases or directories.
- *Collection metadata records* serve both as a general aggregation record for linked items and as an administrative base for managing import and update of items. The aggregations defined by collection metadata records may be 1) administrative, identifying metadata records aggregated by a distinct entity such as a library, 2) semantic, identifying a group of items based on factors such as subject relation, age appropriateness, and the like, or 3) personal, aggregating items based on a person's idiosyncratic preferences. Collection records may describe collections of digital resources for which there is no available item metadata. Furthermore, an item may be aggregated within more than one collection.
- *Link metadata records* describe the relationships between metadata records within the repository. The most common example of a link record describes the parent/child relationship between a collection record and the item records that it aggregates.

Qualified Dublin Core (with DC-ED extensions[1]) is the normalized format for both item and collection metadata. As stated earlier, eight native metadata formats are supported. The MR stores both the native metadata and the DC metadata.

In addition to the descriptive metadata that describe the resource item, metadata records also contain local administrative metadata that describes the metadata record itself. Examples include the source of the metadata, the date the record was last modified, and who has authority to modify the record.

Each metadata record, regardless of type, is identified by a unique NSDL handle, which consists of two parts: a naming authority (which is assigned when a collection is registered with the NSDL), and a unique local name or number assigned by that naming authority. Note that the handle identifies the metadata record, not the item or collection for which it acts as a surrogate. Thus, multiple metadata records about the same content resource have different identifiers.

**Adding Metadata Records to the MR**

As illustrated in Figure 3, all metadata is processed through a "front porch", from which the MR batch harvests the metadata. This intermediate processing stage allows a variety of automatic processing of the metadata including verification for accuracy and integrity, normalization, and cross walking. This ensures that metadata committed to the MR meets a certain level of consistency.

Metadata enters the front porch, and thereby the MR by four mechanisms:

*Metadata ingest via OAI*
The Open Archives Initiative Protocol for Metadata Harvesting [26] defines an HTTP-based mechanism for data providers to expose their metadata for automatic harvesting. Since the protocol mandates that supporting repositories provide metadata in Dublin Core format and also requires a unique identifier for each record, many of the prerequisites for NSDL metadata records will have been met before the metadata is harvested. Support of the OAI harvesting mechanism by as many collections as possible will increase the efficiency of the NSDL.

*Metadata ingest via FTP, e-mail or web-upload (batch)*
Collections may also submit metadata, encoded in one of the eight NSDL-accepted formats, in these encodings: XML-based text file, Excel spreadsheet (.csv files), or tab-delimited text file. These files are then programmatically cross-walked in the repository front porch to be harvested by the repository.

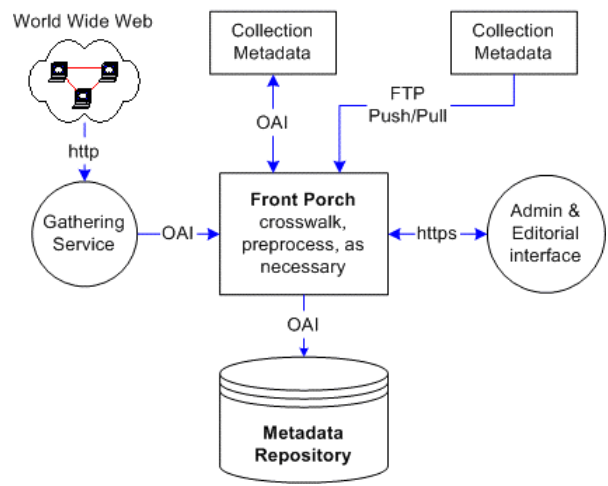

Figure 3 - Ingest into the metadata repository

*Metadata ingest by direct entry*
Collections may authorize a limited number of people to create, edit, and delete Item metadata records. GEMCat [5], from the ERIC Clearinghouse on Information & Technology [4] at Syracuse University, has been selected as an initial user interface for this purpose. This interface provides support for the direct entry of metadata only in Dublin Core format.

*Metadata ingest by gathering*
Metadata may also enter the MR via a combination of web crawling of open-access collections and automatic generation from that content. The Core Integration team at Cornell has been experimenting with Compaq's Mercator [14] extensible web crawler for this process. Automatic generation of metadata from textual content is the subject of another NSDL project at Syracuse University [18].

**Harvesting Metadata from the MR**
Both the Dublin Core and native metadata records in the MR are made available to services through the OAI protocol.  No restriction is placed on access to the Dublin Core records, making it possible for any party, NSDL-funded or external, to create a service building on the data in those records.  In most cases, the contents of the native metadata records are also open-access.  However, experience with the Site for Science prototype has shown that certain collection providers consider their rich metadata to be an important asset and specify that it be made available only to "privileged" consumers, such as the core search and discovery service.

Some of the structural aspects of the MR, external to individual metadata records, have utility for external services.  One example is the linkage relationships among various metadata records.  Since the OAI protocol is ill-suited for exposure of such data, an XML-RPC (XML Remote Procedure Call) interface with data access methods is also provided.

**SEARCH AND DISCOVERY**
The intent of the search and discovery component of the NSDL is to provide fundamental capabilities for locating resources and collections within the library. Search services allow any item represented in the MR to be found, provided that it includes appropriate and accurate information. This service must gracefully deal with the reality that metadata in the MR varies dramatically in its quality and comprehensiveness.

Here we briefly outline the nature of items that are searchable, the capabilities provided by the search and discovery component, and precisely how that component fits into the larger NSDL architecture.

**Searchable Items**

Generally speaking, any item that has metadata in the MR is accessible via the search and discovery component. The only exceptions are those items with metadata consisting only of a resource identifier. Searchable components correspond to all Dublin Core elements, including DC-ED extensions. Since, as described in the previous section, other metadata vocabularies are mapped via the front porch to DC, these searchable components also include information from non-DC vocabularies.

Where possible, the search and discovery component also allows search by the actual content of the resources corresponding to a record in the MR. The content is accessed using open network protocols (e.g., HTTP or FTP) linked via the identifier in the metadata record. In the first phase of the NSDL, for content to be available via search, the content must be freely accessible over the Internet and it must be stored in one of a small set of textual formats: open formats such as ASCII text and html and a handful of proprietary formats such as PostScript, PDF, and Microsoft Word. Eventually we hope to provide content-level search for password-protected or otherwise restricted items. This will be possible as a result of future collaborative work with the rights broker service (described later in this paper). However, this initial limitation simplifies rights management: the only information used by the search service is the publicly available content of the MR and the freely accessibly content of the items represented there.

Content-based search initially supports only textual queries, meaning that non-text items (e.g., images, sound recordings) are accessible via metadata only. Thus, the discoverability of non-textual items depends on adequate descriptive metadata: e.g., in the DC title or description elements.

Since, as described in the previous section, the MR also includes collection-level metadata, information on collections is also indexed and available via the search and discovery service. Eventually, we may if appropriate include (within the searchable aspects of a collection) the information that is collected from the items identified as part of the collection.

**Search Services**

Search and discovery services are available using metadata, content, and any combination of them both. The most likely combination is a content-based search with metadata fields being used to restrict the items returned. For example, "find items that discuss water pollution at the elementary grade level that were published after 1995" includes a content search ("water pollution") and two metadata restrictions ("elementary grade level" from the DC "audience" element and "published after 1995" from the DC "date" element).

The query language that supports these services is independent of the architecture. However, the service provides a language modeled after Z39.50 type 102 ranked list queries [23]. That query language specification was designed to incorporate most features that are common in commercial and web search engines. Our intent is to use the specification as a model, not necessarily as the actual query language.

Type 102 queries include three parts:

- The *restriction set* specifies the set of items over which the query should be run. This includes specification of collections, publication date, author, publisher, reading level, or any other metadata element item that is contained in the repository. No items will appear in the ranked list that does not satisfy this restriction set. Further, the restriction does not in any way indicate the order in which items should be ranked.

- The *ranked query expression* describes the information in which the searcher is interested. It may include free text or Boolean expressions that will be compared against descriptive fields and/or the textual content of items. It may include ranking preferences based on the value of metadata elements (e.g., "all things being equal, give preference to recently published items").

- A *list of known items* specifies items retrieved in the past that are known to be related to the query. It may include indication of which are relevant and which are not, or it may be an earlier query result on which the ranked query expression should be run—i.e., relevance feedback. (Capabilities using this list will not be available in the first implementations of the search and discovery service.)

Type 102 queries contain other capabilities and detailed specifications of a query language. It is not yet clear which (if any) of those will be used.

**Search as a Component**

A goal of the NSDL architecture is to make components as independent of each other's internal representations as possible. To that end, the search and discovery services are implemented as a server that accepts queries and returns ranked lists of items that match. Figure 4 shows the interaction of the search and discovery service with other NSDL components.

The search engine that is being used for the initial version of the NSDL is a commercialized version of InQuery, the research search engine developed at the University of Massachusetts Center for Intelligent Information Retrieval [9, 12]. Because the requirements of the NSDL can only be learned through experience with phase 1, the system is being designed to encompass the possibility that a different search engine will be needed. That is, very few (perhaps no) InQuery-specific capabilities are being incorporated into the query language.

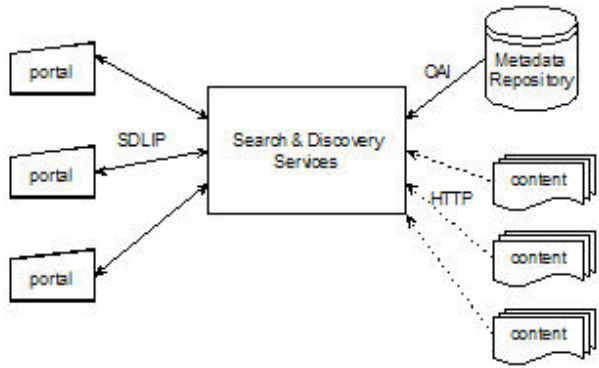

**Figure 4 - Search services interactions with other NDSL components**

The search component acquires its list of searchable items by harvesting the contents of the MR using the OAI protocol [26]. This protocol allows the repository's contents to be acquired initially, and for the search component to update its indexes regularly by harvesting new and changed items. The open protocol interface also means that search interfaces may easily be ported in the future to other databases.

Finally, the search engine interacts with its clients—portals, not people—using the SDLIP protocol [21]. SDLIP defines how queries are sent from client and server, and how results are returned. It does not define the actual query language, or the specifics of the returned list. As discussed above, we are using a query language modeled on Z39.50 Type 102 queries.

The result of a search is a ranked list of items. The portal receives (via SDLIP) some summary information about the items (title, score, etc.) and a pointer to the item. That pointer contains sufficient information to allow the rights broker to determine how the searcher can access the retrieved items, but the search service itself does not in any way provide or enforce access to the content of retrieved items.

## ACCESS MANAGEMENT

NSDL core access management must accommodate a widely varying set of requirements from the users of the library and from the providers of intellectual property, i.e. the NSDL collections, and the providers of other services to the library.

While many items in the library will be freely available and anonymous user access is permitted, it is also expected that access to some materials will be restricted, and that authors and collection publishers will want to know how materials are being used and by what kinds of people.

The NSDL core access management system will balance these different requirements by mediating transactions between the consumers of and providers to the library, acting as a "trusted third party."  Wherever possible, it delegates user authentication and authorization functions to the administrators of the user communities.  It also permits collection managers to specify the terms upon which their materials can be used.  The NSDL community of stake-holders is working together to establish standard ways to formally specify user communities and usage terms [30].

Collection and service projects may then deliver library items and services to individual users without knowing actual user identities. The access management system can provide reports to collection and service providers, showing in detail the nature and demographics of the use of materials and services.

### Authentication and Authorization

The core access management system relies on standard (e.g., Kerberos [25], LDAP [31]) or emerging protocols (e.g., Shibboleth [7]) to distribute identity verification (authentication) and cohort membership (authentication) to the administrators of distinct communities of users.  In other words, the user's "home" institution performs user identity and capability management.

Supported identity services range from those requiring individual logon to group-based mechanisms, such as organizational proxies and network topology identity (IP address), to anonymous identity.

When the home institution does not support standard protocols the NSDL may provide simple gateway services.  When the home institution does not have an existing infrastructure to manage its users, the NSDL provides a simple user registry itself.

### User Profile Server

The NSDL core architecture includes a Profile Server, which holds attributes associated with a user.  Portal interfaces, or other services, may use the Profile Server to store and retrieve information to customize a user's experience.  For example, a portal may store search preferences and histories, disability information, grade level, etc. in the Profile Server and adjust its interface behavior when the user next visits the portal.

The Profile Server engages in authentication and authorization transactions on behalf of its clients, and forwards no individual identity information to them.  It therefore enables services customized for individual users but does not disclose the individuals' identity to the service.

Certain demographic or other aggregate information are stored in the Profile Server, as they are discovered during the authorization process, and are made available to clients.  For example, grade level or instructor status may be obtained from the user's home authorization service and then would normally be offered by the Profile Server.

**Rights Management Broker**
The core architecture also includes a Rights Management Broker service.  The Rights Broker enforces access decisions for items in the library based on the characteristics of the user and of the item.

User information is obtained from authorization systems, as above. Item access information may be specified as part of the metadata record in the MR. Item information may be different for each item in a collection, or groups of items may share the same access terms, or the terms may be the same for entire collections.

The Rights Broker maintains hierarchy information for both users and library items.  For example a junior at Berkeley might also be known as a "higher-ed undergraduate".  The NSDL community is developing standards for appropriate hierarchical categories. Meaningful categories must match those used by collections because the collections specify the access for their items using these categories.

Collections also specify the acceptable type of item use, such as "open", "for personal teaching only", or "for fee". Thus, an item might be characterized as "open access for K-12 students; for fee for corporate research department".

When an item is requested, the Rights Broker does a simple match within user and collection hierarchies to determine if the user may retrieve the item and if a use statement should be attached to the item.

As with the Profile Server, the Rights Broker manages the transaction on behalf of the user so that no individual formation is disclosed to the collection.

**USER INTERFACE AND PORTALS**
Users of the NSDL will access  collections and services through portals.  The users will be very diverse, including students, instructors, the public at all levels, librarians, NSDL federated partners, and community interest groups. Therefore there will be several portals (e.g., the main portal, specialized portals and personalized portals).  Each portal will be an organized, coherent view of resources, an aggregation of choices, a view into the NSDL, a series of web applications, or a system.

The portal architecture and user interface concepts are based on principles developed in the 2000-2001 Site for Science prototype [10]. Key requirement are low-cost and scalability.  Therefore, the visual and functional design elements are primarily driven by data from the MR rather than by "hand-made" content.

The goal is to provide many different views of the library, but with user interfaces that convey the sense of a single library.  Thus the visual design needs to be broad in its appeal, highly intuitive, and customizable.

The same core services will be available through each portal, although the presentation may be different. This leads to the use of non-hierarchical user interface displays, rather than by the traditional "drill down" method featured in many static web sites.

Core services available to all portals are similar to those in more traditional library applications. **Search** services support both simple and advanced forms, producing search results that can include collections, items, and annotations.  **Filtering** allows users to limit the collections searched, formats returned, as well as other parameters.  Search results display with contextual cues relating item records to collections, indicators of the presence of annotations or reviews, and other clues to assist in the determination of relevance of the results to the user.  A **Browse** feature allows users to find appropriate collections to explore or search; the display of collections available includes descriptive text and information about the scope of the collection.  The browse feature allows users to explore collections by topic or by other criteria. **Login** gives users access to a wider variety of options and tools, based on their permissions. A full array of context- sensitive **Help** services is available as well.

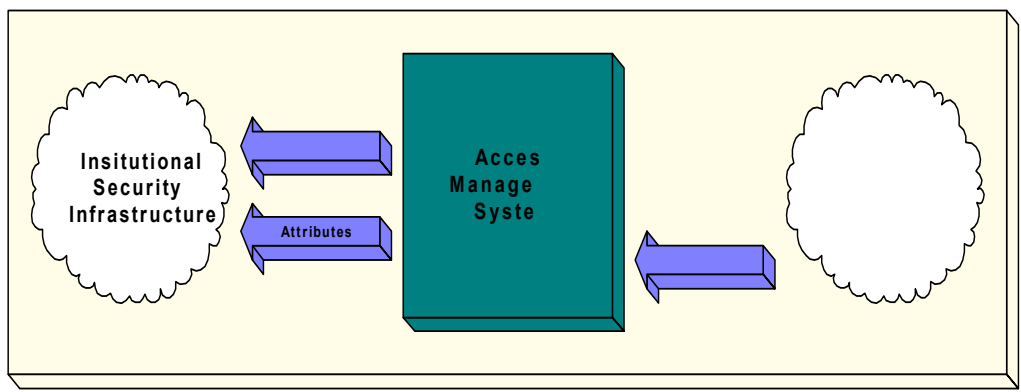

Figure 5 - Components of the Access Management System

**Exhibits** highlighting featured and new collections are available from each portal, based on the special nature of the particular portal (new collections aimed at student users may appear, for instance, in exhibit space in the portals for students and teachers, while collections aimed at teachers may appear only in the latter). **Tools** appropriate for particular user groups are available when visitors select the Tools Portal.

An example of a tool being developed for use by the NSDL community of users and library builders is NewsMaker, designed for gathering, disseminating, and publishing news and events. Users have the ability to search for and redistribute information about their own projects and research by various means: exporting a compiled newsletter to a mail program for targeted redistribution, adding discoverable information to the NSDL catalog, as well as adding content for use in auto-updating displays in NSDL Portals.

Central to the vision of NSDL is the creation of various kinds of portals. The NSDL architecture supports three types in the initial version:

**The Main Portal**—Library visitors access the pre-configured, general use main portal for core library services: login, search, help, browse, my NSDL (explained below), specialized portals, tools, and communication. A corresponding set of administrative tools is available on Main Portal pages to users based on user authorization. Any user, logged in or not, may browse or search the Library Catalog. The same access model holds true for each of the top-level services in the Main Portal. Clear and common language is used throughout, including meaningful labels.

**Specialized Portals**—Registered NSDL community members access specialized portal-building tools that allow them to create and feature content of particular interest to their discipline or community in the NSDL User Interface. Specialized portal services accumulate content managed in a content repository. Metadata is created for all content created as part of a specialized portal. Five specialized portal models are under consideration for the first NSDL release in the fall of 2002: For Students, For Teachers, News, Using Data in the Classroom, and Science Pictures.

**Personalized Portals**—Registered users access several specialized portals, and may personalize some aspects of their experience with NSDL. As demonstrated in the Site for Science Prototype, the ability to create a personalized portal is a highly desirable feature. Other personalization may also be offered, to enhance the ability of users to transfer information to other applications or to store links for later use. A simple version of a personalized portal service, "My NSDL," is planned for the initial release of the NSDL; more sophisticated versions will be offered in later releases. Personalized portal services accumulate content managed in a content repository. Personalized portals may be "published" at the request of the user; this action requires the user to evaluate and approve an automatically created metadata record or records, which then become searchable. Metadata for "published" personalized portals is transferred to the MR and is available to outside services.

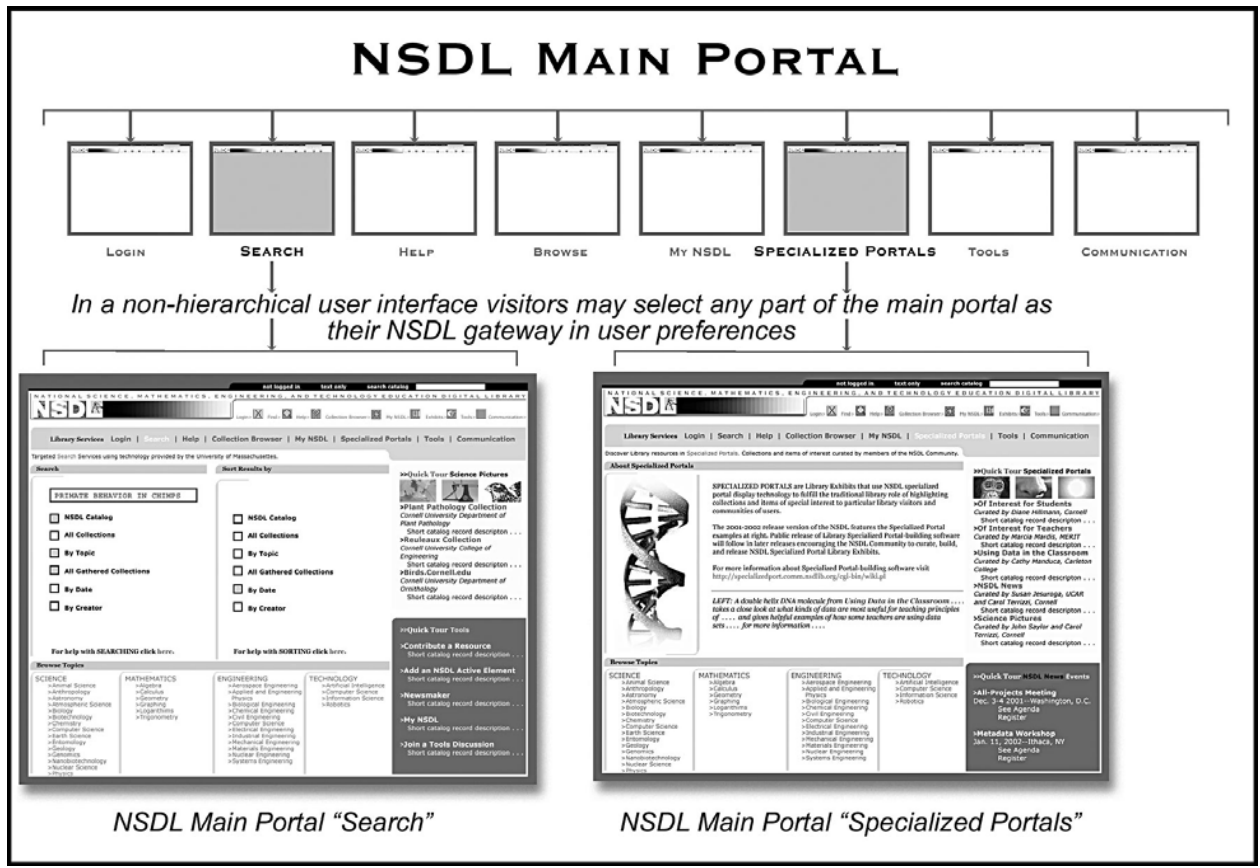

**Annotations and Reviews**

The planned initial release of the NSDL provides users with the ability to contribute textual annotations or reviews of resources. Support for other formats in addition to text will follow in subsequent releases, as will the inclusion of annotations provided by other services. Annotations and reviews, some by casual users and others by educators and other professionals, enhance the value of the library and focus attention on the resources seen as higher in quality or more useful to particular groups of users. For example, a fourth grade teacher using resources on comets for her class could annotate those resources based on her classroom experience, thus providing valuable additional information for other teachers using the same resources.

Annotation records are stored in a content repository, along with the relevant metadata. The metadata, in the case of textual annotations created automatically from the annotation record itself, is harvested by the MR.  As a result, annotations become searchable in the same manner as any resource in the library.   Users may also specify the presence or absence of annotations, as well as information on their source, as criteria for searches in the library.

In the future, metadata on annotations collected by other NSDL services will also be harvested, including such important assets as librarian-contributed annotations on specific resources collected by the "Ask ERIC" service now being expanded by the Virtual Reference Desk Project at Syracuse University [8].

**QUESTIONS AND CONCLUSIONS**

One of the intriguing aspects of digital library research is that grand challenges exist at both the fundamental technology level and at the large-scale integration level. Almost a decade of government and private funding of digital library research projects has produced important results at the fundamental technology level in areas such as web searching, non-textual information retrieval, and knowledge representation.  The successes at the level of large-scale integration are arguably less evident.

NSDL provides the opportunity to produce results in this area.  While the technology presented in this paper may seem more prosaic than that in some other DL research projects, the integration challenges are indeed formidable. A number of outstanding issues will influence the success of the NSDL.

- *Metadata quality:* Unlike the traditional library environment, NSDL metadata originates from sources with varying expertise and following loosely defined and divergent standards.  Can intelligent search algorithms ameliorate and normalize this diverse metadata?

- *Cost*: Related to the previous issue is the actual cost of running a production NSDL.  Can automatic metadata creation and normalization and web crawler-based gathering replace substantial human effort?

- *Integrity and longevity of content*: Libraries deservedly take pride in their selection and preservation tradition. How will the NSDL architecture ensure that associated content has consistent access and reliable integrity and appropriateness for the intended audiences?

- *Intellectual property protection*: Content providers are understandably nervous about the safety of their intellectual property.  Can the NSDL devise the necessary policies and mechanisms that make it possible for the widespread participation of cautious commercial publishers?

- *Richer Functionality*: The intent of the architecture described here is to lay a scalable base for increased functionality in the future.  How will the "spectrum of interoperability" work on a practical level, especially in terms of maintaining a consistent and usable interface for the user?

There are many other substantive questions that the NSDL will face as it grows.  The opportunity to face these questions on a large scale and match them with workable solutions makes the NSDL among the most exciting laboratories for the digital library community.

**ACKNOWLEDGEMENTS**

The authors acknowledge the support of the entire NSDL community, notably Dave Fulker and Lee Zia. This work is funded by the National Science Foundation under grant numbers 0127308 and 0127520. The work at the University of Massachusetts is supported in part by the National Science Foundation Cooperative Agreement number ATM-9732665 through a subcontract from the University Corporation for Atmospheric Research (UCAR).

---

[i] Cornell contributor email addresses: W. Arms (wya@cs.cornell.edu), S. Gan (sgan@cs.cornell.edu), D. Hillmann (dih1@cornell.edu), C. Ingram (cingram@cs.cornell.edu), D. Krafft (dean@cs.cornell.edu), R. Marisa (rjm2@cornell.edu), J. Phipps (jphipps@cs.cornell.edu), J. Saylor (jms1@cornell.edu), C. Terrizzi (clt6@cornell.edu). Columbia contributor email addresses: W. Hoehn (wassa@columbia.edu), D. Millman (dsm@columbia.edu). U. Massachusetts contributor email addresses: J. Allan (allan@cs.mass.edu), S. Guzman-Lara (guzman@cs.umass.edu), Tom Kalt (kalt@cs.umass.edu)